\documentclass[aps,prl,twocolumn,nofootinbib,longbibliography,amsfonts,amssymb,amsmath,superscriptaddress]{revtex4-2}
\usepackage[dvipsnames]{xcolor}
\usepackage{microtype}
\usepackage{IEEEtrantools}
\usepackage{mathrsfs}
\usepackage{amsthm}
\usepackage{enumitem}
\usepackage{varwidth}
\usepackage[normalem]{ulem}
\usepackage{graphicx}
\usepackage{stackengine}
\usepackage{cprotect}
\usepackage[caption=false]{subfig}
\usepackage{multirow}

\usepackage[pdftex]{hyperref}

\usepackage[USenglish]{babel}
\usepackage{microtype}

\usepackage{tikz}
\usetikzlibrary{shapes.geometric, arrows.meta, backgrounds,
  fit, graphs, quotes}

\tikzset{
  node distance=2cm,
  io/.style={trapezium, align=center, rounded corners, trapezium left angle=70,trapezium right angle=-70,minimum height=0.5cm, text centered, draw=black, fill=YellowOrange!20 },
  context/.style={trapezium, align=center, rounded corners, trapezium left angle=70,trapezium right angle=-70,minimum height=0.5cm, text centered, draw=black, fill=Cerulean!20 },
  process/.style={rectangle, align=center, minimum width=.75cm, minimum height=.75cm, text centered, draw=YellowOrange, line width=1.5pt},
  point/.style={circle,inner sep=0pt,minimum size=1pt, fill=black},
  op/.style={circle, minimum size=2pt, inner sep=0pt, text centered, draw=black},
  >={stealth},
  every new ->/.style={thick},
  flow/.style={thick, YellowOrange},
  line/.style={draw,thick,-stealth}
}

\begin{document}

\title{Real-time gravitational-wave science with neural posterior estimation}

\author{Maximilian Dax}
\email{maximilian.dax@tuebingen.mpg.de}
\affiliation{Max Planck Institute for Intelligent Systems, Max-Planck-Ring 4, 72076 T\"ubingen, Germany}
\author{Stephen R. Green}
\email{stephen.green@aei.mpg.de}
\affiliation{Max Planck Institute for Gravitational Physics (Albert Einstein Institute), Am M\"uhlenberg 1, 14476 Potsdam, Germany}
\author{Jonathan Gair}
\email{jonathan.gair@aei.mpg.de}
\affiliation{Max Planck Institute for Gravitational Physics (Albert Einstein Institute), Am M\"uhlenberg 1, 14476 Potsdam, Germany}
\author{Jakob~H.~Macke}
\affiliation{Max Planck Institute for Intelligent Systems,  Max-Planck-Ring 4, 72076 T\"ubingen, Germany}
\affiliation{Machine Learning in Science, University of T\"ubingen, 72076 T\"ubingen, Germany}
\author{Alessandra Buonanno}
\affiliation{Max Planck Institute for Gravitational Physics (Albert Einstein Institute), Am M\"uhlenberg 1, 14476 Potsdam, Germany}
\affiliation{Department of Physics, University of Maryland, College Park, MD 20742, USA}
\author{Bernhard Schölkopf}
\affiliation{Max Planck Institute for Intelligent Systems,  Max-Planck-Ring 4, 72076  T\"ubingen, Germany}

\begin{abstract}

  We demonstrate unprecedented accuracy for rapid gravitational-wave
  parameter estimation with deep learning. Using neural networks as
  surrogates for Bayesian posterior distributions, we analyze eight
  gravitational-wave events from the first LIGO-Virgo
  Gravitational-Wave Transient Catalog and find very close
  quantitative agreement with standard inference codes, but with
  inference times reduced from $O(\text{day})$ to 20 seconds per
  event. Our networks are trained using simulated data, including an
  estimate of the detector-noise characteristics near the event. This
  encodes the signal and noise models within millions of
  neural-network parameters, and enables inference for any observed
  data consistent with the training distribution, accounting for noise
  nonstationarity from event to event. Our algorithm---called
  ``DINGO''---sets a new standard in fast-and-accurate inference of
  physical parameters of detected gravitational-wave events, which
  should enable real-time data analysis without sacrificing
  accuracy.

\end{abstract}

\maketitle

\emph{Introduction.---}Since the first detection of a signal from a
pair of merging black holes~\cite{Abbott:2016blz}, gravitational waves
have quickly emerged as an important new probe of gravitational
theory~\cite{LIGOScientific:2020tif}, neutron-star
physics~\cite{Abbott:2018exr}, cosmology~\cite{Abbott:2019yzh}, and
black-hole astrophysics~\cite{Abbott:2020gyp}. These scientific
successes were made possible by a growing rate of detections by the
LIGO~\cite{TheLIGOScientific:2014jea} and
Virgo~\cite{TheVirgo:2014hva} observatories, and their subsequent
analysis and characterization as signals from merging compact binary
systems. The LIGO and Virgo Collaborations (LVC) have now published
results from over 50 such
systems~\cite{LIGOScientific:2018mvr,LIGOScientific:2020ibl}, and this number
promises to grow ever-faster as detectors are made more sensitive in
the future~\cite{Abbott:2020qfu}.

Given a detection, Bayesian inference is used to characterize the
originating source~\cite{LIGOScientific:2019hgc}. This is based on
having models for the signals and the detector noise. For
gravitational waves, signal models take the form of waveform
predictions $h(\theta)$ depending on the source parameters $\theta$
(masses, location, etc.). Waveform models are based on solutions to
Einstein's equations (and any relevant matter equations) for the
two-body dynamics and gravitational radiation, using a combination of
numerical-relativity and perturbative
calculations~\cite{Buonanno:1998gg,Bohe:2016gbl,Varma:2018mmi} and
phenomenological
fitting~\cite{Purrer:2015tud,Khan:2015jqa,Varma:2018mmi}. Detector
noise is typically modeled as stationary and Gaussian, with some
spectrum which can be estimated empirically. Together, these
``forward'' models give rise to the likelihood $p(d|\theta)$ for the
observed \emph{strain} data $d$, which is assumed to consist of a
signal plus noise. With the choice of a prior $p(\theta)$ over
parameters, the posterior distribution is given via Bayes' theorem,
\begin{equation}\label{eq:Bayes}
  p(\theta|d) = \frac{p(d|\theta)p(\theta)}{p(d)},
\end{equation}
where $p(d)$ is a normalizing factor called the evidence. The
posterior gives our belief about the source parameters, given the
observed data.

The task of inference is to characterize the posterior by drawing
\emph{samples} from it. This can be accomplished with stochastic
algorithms like Markov chain Monte Carlo (MCMC). The LVC have
developed software tools such as LALInference~\cite{Veitch:2014wba}
and Bilby~\cite{Ashton:2018jfp,Romero-Shaw:2020owr,Speagle_2020} to
carry this out.  However, these algorithms are computationally
expensive as they require many likelihood evaluations for each
independent posterior sample $\theta\sim p(\theta|d)$, and each
likelihood requires a waveform simulation. An analysis producing
$\sim 10^4$ independent samples typically requires millions of
waveform evaluations and a total inference time of hours to months,
depending on the signal duration and waveform model. More
physically-realistic waveform models~\cite{Ossokine:2020kjp} are also
more costly, so carrying out inference for all events with the best
models is an enormous computational effort. When rapid results are
desired---for alerts to trigger electromagnetic follow-up of transient
phenomena~\cite{GBM:2017lvd}, or when processing large numbers of
events---accuracy usually has to be traded off for speed, by restricting to a limited set of fast models~\cite{Pratten:2020fqn,Pratten:2020ceb} or
specialized inference
algorithms~\cite{Singer:2015ema,Lange:2018pyp,Cornish:2021wxy}.

In this Letter, we describe an alternative approach to
gravitational-wave inference which delivers both dramatically reduced
analysis time \emph{and} high accuracy, in stark contrast to the
trade-off intrinsic to standard algorithms. The basic idea is to
produce a large number of simulated data sets (with associated
parameters), and use these to train a type of neural network known as
a \emph{normalizing flow} to approximate the posterior. The trained
network can then generate new posterior samples extremely quickly once
a detection is made. This bypasses the need to generate waveforms at
inference time, thereby \emph{amortizing} the expensive training costs
over all future detections. The general approach of building such
``surrogate'' inverse models is called \emph{neural posterior
  estimation}
(NPE)~\cite{papamakarios2016fast,lueckmann2017flexible,greenberg2019automatic},
and is beginning to see application in several scientific
domains~\cite{Cranmer:2019eaq}. When applied to gravitational waves,
with all of the optimizations we describe, we call the method
\emph{Deep INference for Gravitational-wave Observations}, or
\emph{DINGO}.

NPE and conventional methods both involve the same inputs: a prior and
a likelihood.  A key difference, however, is the way in which the
likelihood is used: for conventional methods, its density is
\emph{evaluated}, whereas for NPE it is used to \emph{simulate data},
i.e., $d\sim p(d|\theta)$. This distinction is important
when dealing with nonstationary or non-Gaussian detector noise, for
which an analytic likelihood is either expensive or unavailable. In
this case, one could nevertheless simulate data, in a noise-model-independent
way, by injecting simulated signals into real noise. Our present focus is on speed and on validating DINGO on real data with the common assumption of stationary-Gaussian noise, but
the ultimate aim of more accurate inference using real noise should be
kept in mind.

There have been several previous studies that applied NPE or related
approaches to gravitational
waves~\cite{Gabbard:2019rde,Chua:2019wwt,Chatterjee:2019gqr,Green:2020hst,Green:2020dnx,Delaunoy:2020zcu,Krastev:2020skk,Shen:2019vep};
see also~\cite{Cuoco:2020ogp}. However, most of these are limited in
some way: they either restrict the number of parameters or the
distributional form of the posterior, or they do not analyze real
data, or there are clear deviations from results obtained using
standard algorithms. The best performance to-date was achieved in the
study~\cite{Green:2020dnx} by some of us. This was the only study to
infer all 15 parameters\footnote{Parameters consist of detector-frame
  component masses $(m_1, m_2)$, time of coalescence at geocenter
  $t_c$, reference phase $\phi_c$, sky position $(\alpha, \delta)$,
  luminosity distance $d_L$, inclination angle $\theta_{JN}$, spin
  magnitudes $(a_1, a_2)$, spin angles
  $(\theta_1, \theta_2, \phi_{12}, \phi_{JL})$~\cite{Farr:2014qka},
  and polarization angle $\psi$.} of a binary black hole (BBH) system
in real data and demonstrate close agreement to standard
samplers. However, even that study did not achieve full amortization,
as it did not address the fact that detector noise varies from event
to event. Rather, the neural network of~\cite{Green:2020dnx} was tuned
to the noise power spectral densities (PSDs) of the detectors at the
time of the event analyzed, and it would require retraining for each
new event.

We now present for the first time completely amortized inference for
BBHs using DINGO. This is achieved by \emph{conditioning} the neural
network not only on the event strain data, but also on the detector
noise PSD, which can be estimated using nearby
data~\cite{Veitch:2014wba}. We also achieve unprecedented accuracy
thanks to a new iterative algorithm for time-shifting the coalescence
times, as well as various architecture improvements. We use our
trained networks to analyze all events in the first Gravitational-Wave
Transient Catalog (GWTC-1)~\cite{LIGOScientific:2018mvr} with
component masses greater than $10~\mathrm{M}_\odot$ (our prior bound)
and find close (sometimes indistinguishable) quantitative agreement
with standard algorithms. This Letter sets a new standard for rapid
gravitational-wave inference, which should enable real-time
gravitational-wave science in the near future. It shows that NPE has
moved beyond toy models and is competitive with conventional
algorithms. More broadly, it provides a demonstration of these new
methods in a realistic use case, which we hope will inspire wider
adoption in experimental science.

\begin{figure}
  \centering
    \includegraphics[width=0.47\textwidth]{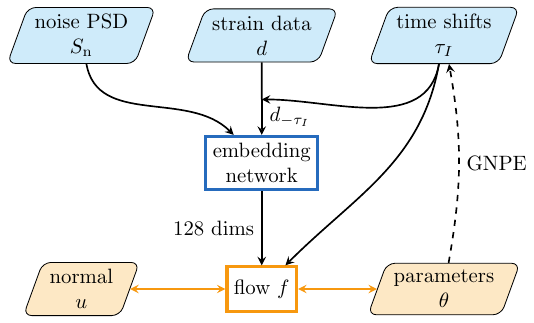}
  \caption{DINGO flow chart. The posterior distribution is represented
    in terms of an invertible \emph{normalizing flow} (orange), taking
    normally-distributed random variables $u$ into posterior samples
    $\theta$. The flow itself depends on a (compressed) representation
    of the noise properties $S_{\text{n}}$ and the data $d$, as well
    as an estimate $\tau_I$ of the coalescence time in each detector
    $I$. The data are time-shifted by $\tau_I$ to simplify the
    representation. For inference, the iterative \emph{group equivariant neural posterior estimation} (GNPE) algorithm is
    used to provide an estimate of $\tau_I$, as described in the main
    text.}
  \label{fig:NDE}
\end{figure}

\emph{Method.---}The central object of DINGO is the density-estimation
neural network, which defines a conditional probability distribution
$q(\theta|d)$. This should be distinguished from the posterior
$p(\theta|d)$, which $q(\theta|d)$ learns to approximate through
training. We use so-called normalizing
flows~\cite{rezende2015variational,kingma2016improved,papamakarios2017masked}
to define a sufficiently flexible $q(\theta|d)$ via a $d$-dependent
mapping $f_d:u\mapsto\theta$ from a simple ``base'' distribution
$\pi(u)$,
\begin{equation}
  q(\theta|d) = \pi\left(f_d^{-1}(\theta)\right)\left|\det J_{f_d^{-1}}\right|.
\end{equation}
If $\pi(u)$ can be rapidly evaluated and sampled from, and if $f_d$ is
invertible and has simple Jacobian determinant, then $q(\theta|d)$ can
also be rapidly evaluated and sampled from. Following \cite{Green:2020dnx},
we take $\pi(u)$ to be multivariate standard normal, and $f_d$ a
composition of spline coupling flows~\cite{durkan2019neural}, each of
which is defined with a neural network.

The overall structure of DINGO is illustrated in
Fig.~\ref{fig:NDE}. This contains three key enhancements compared to
the study~\cite{Green:2020dnx}. First, since the data generation
process depends on the detector noise PSD $S_{\text{n}}$, we include
this as additional context to the neural network, i.e.,
$q(\theta|d,S_{\text{n}})$. This allows us to tune the network at
inference time to the PSD estimated just prior to the event,
corresponding to standard ``off-source'' noise
estimation~\cite{Veitch:2014wba}.  An alternative would be to estimate
the noise ``on-source''~\cite{Littenberg:2014oda}, but since we
consider only short-duration BBH events here, the off-source approach
is sufficient.

The second enhancement addresses the problem of high-dimensional
observed data by using an additional neural network to first compress
to a small number of features. This network (called an ``embedding
network'') is trained alongside the flow network. Our data is in the
frequency domain, between 20~Hz and 1024~Hz, with 0.125~Hz resolution,
so combined with the PSDs, this gives 24,096 input dimensions for each
of the two or three interferometers. The first stage of the embedding
network maps this linearly to 400 components per detector. To provide
an inductive bias to extract signal information, we seed this layer
with the principal components of clean waveforms from our training
set, and then allow these parameters to float during
training. Following this, a fully-connected residual
network~\cite{he2015deep} compresses to 128 features, which are
provided to the flow.

Finally, we developed a new method to treat time translations of the
strain data. For standard algorithms, inference of
$(\alpha,\delta,t_c)$ requires sampling over waveforms with varying
coalescence times $t_I$ in each detector $I$. Likewise for NPE, the
network must learn to interpret strain data with different $t_I$. For
frequency-domain data, however, time translations correspond to local
phase shifts, which, although explicitly known, are challenging for
neural networks to learn. Indeed, this occupied much of the network
capacity in Ref.~\cite{Green:2020dnx}. Our new
approach---called \emph{group equivariant} neural
posterior estimation (GNPE)---leverages explicit knowledge of the
time-translation symmetry along with \emph{approximate} knowledge of
$t_I$ to simplify the data representation and allow the network to
focus on more nontrivial parameters.  For further details see~\cite{Dax:2021myb}.

For GNPE, we train the network to infer $\theta$ given perturbed
coalescence times $\tau_I$ and manually-time-shifted strain data
$d_{-\tau_I}$. Using maximum likelihood
estimation~\cite{Goodfellow-et-al-2016}, this means we minimize the
loss function
\begin{align}
  \label{eq:loss}
  L ={}& \mathbb{E}_{p(\theta)}\mathbb{E}_{p(S_{\text{n}})}\mathbb{E}_{p(d|\theta,S_{\text{n}})}\mathbb{E}_{\kappa(\delta t_I)}\left[\right.\nonumber\\
  &\qquad\qquad\left.-\log q(\theta|d_{-t_I(\theta)-\delta t_I},S_{\text{n}}, t_I(\theta)+\delta t_I)\right],
\end{align}
with respect to the network parameters \cite{Kingma:2014vow}. Here, $\mathbb{E}$ refers to the expected value over the
specified distribution, which is evaluated stochastically using Monte
Carlo draws. $\kappa(\delta t_I)$ is a uniform
kernel used to perturb $t_I$. For inference, even though we do not
have direct access to $t_I$, all parameters can be inferred using
Gibbs sampling starting with an approximate $t_I$ (obtained, e.g.,
using standard NPE): first, convolve $t_I$ with $\kappa(\delta t_I)$
to obtain $\tau_I$, then use the network to infer a new estimate for
$t_I$; then convolve again and repeat. We find that this converges
after $O(10)$ iterations.

Evaluating \eqref{eq:loss} requires sampling
$\theta^{(i)}\sim p(\theta)$ and
$S_{\mathrm{n}}^{(i)}\sim p(S_{\text{n}})$, and then simulating data
$d^{(i)} \sim p(d|\theta^{(i)},S_{\mathrm{n}}^{(i)})$. Aside from the
PSD sampling, this follows Ref.~\cite{Green:2020dnx} very closely. In
particular we use the same prior over parameters, with
$m_1, m_2 \in [10,80]~\mathrm{M}_\odot$. We train separate networks
for the noise distributions in the first (O1) and second (O2)
observing runs of LIGO and Virgo, with PSD samples estimated
empirically from stretches of interferometer noise
data~\cite{Abbott:2019ebz}. For O1, we choose the distance prior
$[100,2000]$~Mpc. For O2, we train one network for loud events with
distance prior $[100,2000]$~Mpc and another for quieter events with
$[100,6000]$~Mpc. In addition to these two-detector networks, we train
a three-detector network with distance prior $[100,1000]$~Mpc to
analyze GW170814. With future enhancements of network architecture we
expect to cover the entire distance range with a single network.
Finally, as in Ref.~\cite{Green:2020dnx}, training data are generated
from a fixed set of spin-precessing frequency-domain waveforms,
described by the
IMRPhenomPv2~\cite{Hannam:2013oca,Khan:2015jqa,Bohe:2016} model, but
with extrinsic parameters and noise realizations drawn randomly during
training. With training sets of $5\times 10^6$ waveforms, there is no
indication of overfitting. Training takes roughly 10 days on a
single NVIDIA A100.  Further details on the networks and
training are provided in the Supplementary Material.

\begin{figure}
  \includegraphics[width=\linewidth]{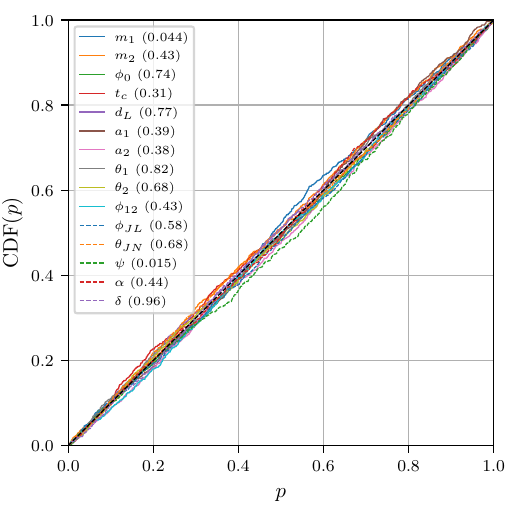}
  \caption{P--P plot for $1000$ injections. The legend shows the $p$-values of the individual parameters, with a combined $p$-value of 0.46.}
  \label{fig:pp}
\end{figure}

\emph{Results.---}As a first test, we evaluate DINGO on data entirely
consistent with the training distribution, i.e., simulated waveforms
in stationary-Gaussian noise. This is an easier task than using
observational data, which includes real signals in noise that is
neither strictly stationary nor Gaussian, and therefore lies outside
the training distribution. We sample posteriors from 1000 simulated
data sets and construct a P--P plot (see Fig.~\ref{fig:pp}). For each
parameter, we compute the percentile score of the true value within
its marginalized posterior, and then we plot the cumulative
distribution function (CDF) of these scores. For true posteriors, the
percentiles should be uniformly distributed, so the CDF should be
diagonal. Kolmogorov-Smirnov test $p$-values are indicated in the
legend, with combined $p$-value of 0.46. This shows that DINGO is
performing properly on simulated data. 

\begin{figure}
  \subfloat[][\label{fig:masses}]{\includegraphics[width=\linewidth]{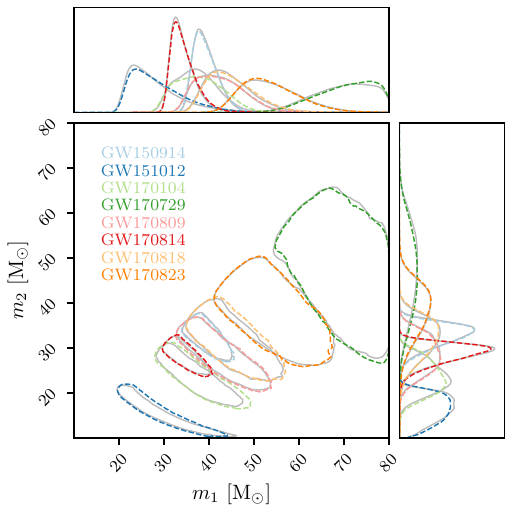}}

  \subfloat[][\label{fig:skymap}]{\includegraphics[width=\linewidth]{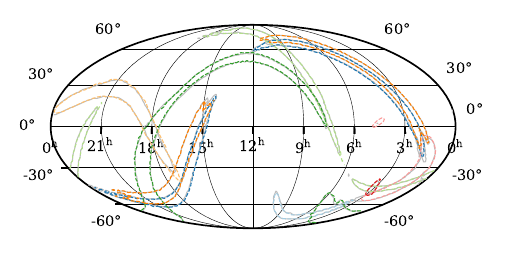}}
  \caption{Comparison of (a) detector-frame component mass and (b)
    sky position posteriors from DINGO (colored) and
    LALInference (gray) for eight GWTC-1 events. 90\% credible
    regions shown.}
  \label{fig:comparison}
\end{figure}

\begin{figure*}
  \includegraphics[width=\textwidth]{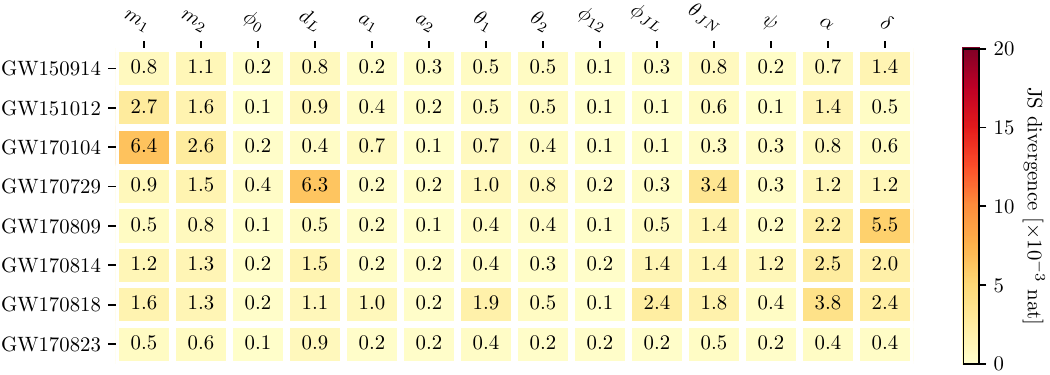}
  \caption{\label{fig:js-heatmap}JSDs between DINGO and
    LALInference marginalized posteriors, averaged
    over 100 realizations. The mean JSD across all
    events and parameters is 0.0009 nat.}
\end{figure*}

We now proceed to our main result, which is a demonstration of
performance on real events. We perform inference on the eight GWTC-1
BBH events compatible with our prior, using both DINGO and
LALInference MCMC. For DINGO, generation of 50,000 sample points with
30 GNPE iterations takes roughly 20 seconds. Comparisons of inferred
component masses and sky position for all events show good agreement
(see Fig.~\ref{fig:comparison}), including multimodality for the sky
position. The one exception is GW170104, where the mass posterior is
slightly flatter.  Nevertheless, 90\% credible intervals are in good
agreement.
      
For quantitative comparisons, we compute the Jensen-Shannon divergence
(JSD) ~\cite{lin1991divergence} between DINGO and LALInference
one-dimensional marginalized posteriors (see
Fig.~\ref{fig:js-heatmap}). This is a symmetric divergence that
measures the difference between two probability distributions, with
values ranging from 0 to $\ln(2)\approx0.69$~nat. We find a mean JSD
across all events and parameters of 0.0009~nat, which is slightly
higher than the variation (0.0007~nat) found between LALInference runs
with identical settings but different random
seeds~\cite{Romero-Shaw:2020owr}. By comparing such LALInference runs,
Ref.~\cite{Romero-Shaw:2020owr} also established a maximum JSD of
0.002~nat for indistinguishability; our results are approaching this
threshold, with two events below for all parameters, and the others
with one to three parameters above. The slight visible disagreement
between mass posteriors for GW170104 is also reflected in larger
JSDs. For comparison, we note that PSD variations (see Supplemental
Material) and the choice of waveform model~\cite{Romero-Shaw:2020owr}
both impact the JSD at a much higher level (0.02~nat). Additional
comparisons between samplers, including posteriors for all events, are
provided in the Supplemental Material.

\emph{Conclusions.---}In this Letter, we introduced DINGO and applied
it to perform extremely fast Bayesian parameter inference for
gravitational waves observed by the LIGO and Virgo detectors. We
analyzed eight GWTC-1 events, and showed excellent agreement with
standard algorithms, with inference times reduced by factors of
$10^3$--$10^4$. This was achieved by conditioning on the detector
noise characteristics and making a number of architecture and
algorithm improvements. The DINGO code is available at 
\href{https://github.com/dingo-gw/dingo}{https://github.com/dingo-gw/dingo}.

A critical component of DINGO is a new iterative algorithm---GNPE---to
partially off-load the modeling of time translations from the neural
network. Although convergence of GNPE may take 20 seconds, initial
samples with slightly reduced accuracy can, however, be produced in
just a few seconds by taking fewer iterations.

Going forward, the next steps are to extend the prior to include
longer-duration binary neutron star
signals~\cite{TheLIGOScientific:2017qsa} (for which rapid results are
especially important to identify electromagnetic counterparts) and to
extend to more physically-realistic waveform models, which include
higher multipole modes and more accurate spin-precession
effects~\cite{Ossokine:2020kjp}. Long and complex waveforms are much
more expensive for standard algorithms, so the relative improvement in
performance should be even more significant. If successful, this would
also enable the routine use of the most physically-realistic
waveforms, resulting in consistently reduced systematic errors. These
extensions will likely require somewhat larger networks and improved
data representation or compression.\footnote{Initial estimates based
  on a singular value decomposition~\cite{Cannon:2010qh} indicate that to accurately
  represent SEOBNRv4PHM BBH waveforms~\cite{Ossokine:2020kjp}, the
  initial layers of our embedding network should be widened by a
  factor of roughly four. Assuming the same number of iterations and
  fixed hardware, the total training time would increase by about 35\%. 
  For binary neutron stars, adopting a frequency-dependent
  resolution~\cite{Vinciguerra:2017ngf,Cannon:2011vi} would limit the expansion of
  the number of frequency bins to a factor of three.}

Another natural extension would be to study signals without making the common stationary-Gaussian idealization for the detector noise during the training stage. 
For DINGO, performing inference with realistic noise is simply a matter of
training with simulated signals injected into real noise realizations
taken from detectors. Using real noise should lead to improved
accuracy that is not possible using standard likelihood-based methods,
and would serve as an excellent demonstration of the advantages of
NPE. For real-time analysis, it will also be necessary to develop
approaches to progressively retrain networks to keep pace with
changing data distributions during an observing run, e.g., as detector
sensitivity is improved. All of these enhancements, particularly the
treatment of nonstationary noise, will be critical for extensions to
future observatories such as LISA.

Deep-learning tools are now ready to analyze the vast majority of
LIGO/Virgo events. In the past, the primary challenge has been in
obtaining sufficiently accurate results, but with DINGO, we have now
achieved this in a realistic context. Through planned future
extensions, we expect that DINGO could become one of the leading
approaches to gravitational-wave inference.

\begin{acknowledgments}
\emph{Acknowledgments.---}We thank S. Ossokine, M. P\"urrer,
C. Simpson and P. Z\"uge for helpful discussions. This research has
made use of data, software and/or web tools obtained from the
Gravitational Wave Open Science Center
(https://www.gw-openscience.org/ ), a service of LIGO Laboratory, the
LIGO Scientific Collaboration and the Virgo Collaboration. LIGO
Laboratory and Advanced LIGO are funded by the United States National
Science Foundation (NSF) as well as the Science and Technology
Facilities Council (STFC) of the United Kingdom, the
Max-Planck-Society (MPS), and the State of Niedersachsen/Germany for
support of the construction of Advanced LIGO and construction and
operation of the GEO600 detector. Additional support for Advanced LIGO
was provided by the Australian Research Council. Virgo is funded,
through the European Gravitational Observatory (EGO), by the French
Centre National de Recherche Scientifique (CNRS), the Italian Istituto
Nazionale di Fisica Nucleare (INFN) and the Dutch Nikhef, with
contributions by institutions from Belgium, Germany, Greece, Hungary,
Ireland, Japan, Monaco, Poland, Portugal, Spain. This material is
based upon work supported by NSF’s LIGO Laboratory which is a major
facility fully funded by the National Science Foundation. M.D. thanks
the Hector Fellow Academy for support. J.H.M. and B.S. are members of
the MLCoE, EXC number 2064/1 – Project number 390727645.  We use
\verb|PyTorch|~\cite{NEURIPS2019_9015} and \verb|nflows|~\cite{nflows}
for the implementation of our neural networks. The plots are generated
with \verb|matplotlib|~\cite{Hunter:2007},
\verb|ChainConsumer|~\cite{Hinton2016} and
\verb|ligo.skymap|~\cite{Singer:ligoskymap}.
\end{acknowledgments}

\bibliography{mybib.bib}

\clearpage
\begin{center}
  \large
  \textbf{Supplemental Material}
\end{center}
\section{Training data}

We perform inference over the full 15D parameter space for
quasicircular binary black holes, which includes detector-frame
component masses $m_1$, $m_2$, time of coalescence at geocenter $t_c$,
reference phase $\phi_c$, sky position (right ascension $\alpha$ and
declination $\delta$), luminosity distance $d_L$, inclination angle
$\theta_{JN}$, spin magnitudes $a_1$, $a_2$, tilt angles $\theta_1$,
$\theta_2$, other spin angles $\phi_{12}$,
$\phi_{JL}$~\cite{Farr:2014qka}, and polarization angle $\psi$. Priors
are taken to be the same as in Ref.~\cite{Green:2020dnx}: standard
over all angles, and uniform in all other parameters, with
$m_1 \ge m_2$, $m_1, m_2 \in [10,80]~\text{M}_\odot$,
$a_1, a_2 \in [0,0.88]$, and $t_c \in [-0.1, 0.1]~\text{s}$. We found
that it was difficult to train a neural network for accurate inference
over the entire relevant range of luminosity distance, so we partition
the prior as shown in Tab.~\ref{tab:networks}.  To perturb the signal
coalescence times $t_I$ for GNPE ,we use a uniform kernel
$\kappa(\delta t_I)$ in the range $[-1, 1]~\text{ms}$.

\begin{table}[b]
  \begin{ruledtabular}
    \begin{tabular}{ccc}
      Observing run & Detectors  & Distance range [Mpc]\\
      \hline
      O1 & HL & [100, 2000] \\
      \hline
      \multirow{3}{*}{O2} & \multirow{2}{*}{HL} & [100, 2000] \\
                    && [100, 6000] \\
      \cline{2-3}
      & HLV & [100, 1000]
    \end{tabular}
    \caption{Neural networks are trained based on noise from a particular observing run, number of detectors, and distance range.}
    \label{tab:networks}
  \end{ruledtabular}
\end{table}

Training data consist of labeled strain data sets $(\theta, d)$ and
associated noise power spectral densities $S_{\text{n}}$. To construct
the data sets, we first draw samples from the prior,
$\theta \sim p(\theta)$. This is done in two stages as in
Ref.~\cite{Green:2020dnx}: first, intrinsic parameters are sampled
ahead of training, and waveforms are generated and saved based on
these; second, extrinsic parameters are sampled during training and
applied to the waveforms, since this involves simple
transformations. For our purposes, intrinsic parameters consist of
$\theta_{\text{intrinsic}} = (m_1, m_2, \phi_c, \theta_{JN}, a_1, a_2,
\theta_1, \theta_2, \phi_{12}, \phi_{JL})$ and extrinsic parameters
are $\theta_{\text{extrinsic}} = (t_c, \alpha, \delta, d_L, \psi)$.

Each strain data set $i$ consists of a waveform with additive
stationary Gaussian noise, $d^{(i)} = h(\theta^{(i)}) + n^{(i)}$. This
is represented in frequency domain, with
$f_{\text{min}}=20~\text{Hz}$, $f_{\text{max}}=1024~\text{Hz}$, and
$\Delta f = 0.125~\text{Hz}$, corresponding to a duration of
8~s. Waveforms are generated using the IMRPhenomPv2 frequency-domain
model~\cite{Hannam:2013oca,Khan:2015jqa,Bohe:2016}, which is fast (so
that comparisons against standard samplers are feasible) and also
includes spin-precession effects. We save intrinsic waveforms to disk
in an SVD representation, which is accurate to a mismatch of
$2\cdot 10^{-5}$ for the $99.9$th percentile of the data. Noise
realizations are generated during training, after first sampling an
associated PSD, i.e., $S_{\text{n}}^{(i)} \sim p(S_{\text{n}})$,
$n^{(i)} \sim p(S_{\text{n}}^{(i)})$. We construct training sets based
on $5 \times 10^6$ sets of intrinsic parameters, but by sampling
extrinsic parameters and noise realizations during training, the
effective size of the training set is infinite in these dimensions.

To construct the empirical PSD distributions $p(S_{\text{n}})$, we
estimate PSDs using noise data from each observing run. Using
\verb|BURST_CAT2| data from GWOSC~\cite{Abbott:2019ebz}, we identify
stretches of at least 1024~s in duration that do not overlap with
events. Each PSD is estimated by taking a 1024~s stretch of data,
dividing this into non-overlapping 8~s subintervals, and using the
Welch ``median-average'' method to average PSDs estimated on each of
these~\cite{Veitch:2014wba}. We use a Tukey window with a roll-off of
0.4~s. For inference, we use the same construction to estimate the PSD
from detector data just prior to an event. The number of PSDs obtained
for each observing run is given in Tab.~\ref{tab:PSDs}.
\begin{table}
  \begin{ruledtabular}
    \begin{tabular}{ccc}
      Observing run&Detector&Number of PSDs\\
      \hline
      \multirow{2}{*}{O1} & H & 2444 \\
                   &L& 2414\\
      \hline
      \multirow{3}{*}{O2} & H & 4670\\
                   &L& 3873\\
      &V&864
    \end{tabular}
    \caption{Number of PSDs estimated for each observing run and detector.}
    \label{tab:PSDs}
  \end{ruledtabular}
\end{table}

\section{Neural network}

There are two main components to our conditional density-estimation
neural network, the embedding network for compressing data to a
sufficiently small number of features, and the normalizing flow, which
produces the Bayesian posterior from these features.

\subsection{Embedding network}

For each of the two or three detectors, the embedding network takes as
input the real and imaginary parts of the whitened frequency-domain
strain data, as well as the inverse amplitude spectral density
(ASD). This results in 24,096 inputs per detector. We provide the
inverse ASD rather than the PSD because of the more numerically stable
behavior of spectral lines. We scale the ASD with a constant factor of
$10^{23}$.

The first embedding layer is a linear mapping that serves to
drastically reduce the number of dimensions. This is initialized based
on a singular value decomposition (SVD) to provide an inductive bias
to facilitate training. 
The strain data are initially projected onto the first $n_\text{SVD}=200$ (complex) singular vectors, defined based on a set of 50,000 signal waveforms drawn from the prior. The inverse ASDs are likewise mapped to $n_\text{SVD}$ complex numbers, which are added to the projected strain; this projection is initialized to 0. 
After this initial layer, the data has therefore
been projected onto $2n_{\text{detectors}}n_{\text{SVD}}$ (real) features
(i.e., either 800 or 1200 components, depending on the number of
detectors). There is no nonlinear activation following this layer.

Following the SVD layer is a fully-connected residual network. This
consists of a sequence of two-layer residual blocks; prior to each
linear mapping are batch normalization layers and ELU activation
functions. We take six blocks each of 1024, 512, 256, and 128 hidden
dimensions, resulting in a total of 48 hidden layers in the residual
network.

\subsection{Normalizing flow}

Our normalizing flow is very similar to that
of~\cite{Green:2020dnx}. This consists of a sequence of coupling
transforms, each of which transforms half of the components of a
sample element-wise based on the context information and the values of
the untransformed components. We use a rational-quadratic spline
coupling transform~\cite{durkan2019neural}, with the same parameters
as in~\cite{Green:2020dnx}, except for the number of residual blocks,
which is reduced from 10 to 5. Between the coupling transforms, the
ordering of the sample components is randomized to ensure all
components are sufficiently transformed by the sequence of
transforms. We also increase the total number of coupling transforms
to 30 from 15 in~\cite{Green:2020dnx}. The flow therefore consists of
300 hidden layers.  In total, the embedding network and the flow
combined have $1.31\cdot 10^8$ learnable parameters for
$n_{\text{detectors}}=2$ and $1.42\cdot 10^8$ for
$n_{\text{detectors}}=3$.

The context information for the flow consists of the 128 features
output from the embedding network, as well as the two or three
perturbed detector coalescence times $\tau_I$.

We also train a neural network to provide an initial estimate for
$t_I$, which is used as a starting point for the iterative GNPE
algorithm. This does not require $\tau_I$ as context, but otherwise
has the same form as the main network.

\subsection{Training}

\begin{figure}
  \includegraphics[width=\linewidth]{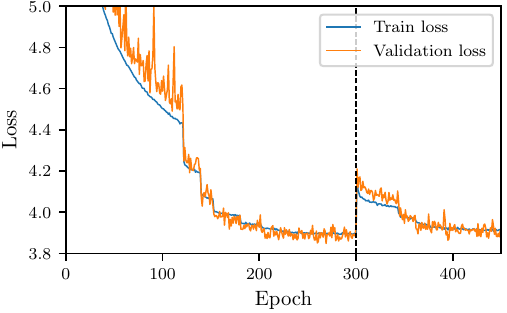}
  \caption{Loss as a function of training epoch for the O1 neural
    network. The vertical line denotes the beginning of the
    fine-tuning period.}
  \label{fig:history}
\end{figure}

Training consists of two stages, an initial pretraining stage and a
fine-tuning stage. During the pretraining stage, the SVD layer is
frozen, and all noise realizations are drawn from an average PSD for
the observing run and detector. This is designed to simplify
early-stage training. During the fine-tuning stage, the SVD layer is
unfrozen, and PSDs are randomly drawn from the empirical distribution,
and noise realizations are drawn from these.

Our networks are trained for 300 epochs of pretraining and 150 epochs
of fine tuning with a batch size of $4096$. We use the Adam
optimizer~\cite{Kingma:2014vow}. We begin training with learning rates
of $3\cdot 10^{-4}$ and $3\cdot 10^{-5}$ in the pretraining and
finetuning stages, respectively, and decrease the learning rate by a
factor of $2$ when the validation loss has not improved in the
previous 10 epochs. For three detectors, we reduce the batch size to
$2048$ and the initial learning rates to $2\cdot 10^{-4}$ and
$2\cdot 10^{-5}$ due to memory limitations.  As shown in
Fig.~\ref{fig:history}, the loss jumps at the beginning of the
fine-tuning stage. This occurs because the distribution of training
data becomes much broader with the inclusion of varying noise
PSDs. The final loss at the end of fine tuning is just above the
pretraining loss, which indicates that the network has learned to
process the varying PSDs to the same performance level of the fixed
pretraining PSD. During training, we reserve 2\% of the training set
for validation to check for overfitting. Since the training and
validation loss are in close agreement in Fig.~\ref{fig:history} we
conclude that overfitting is minimal.  Training 450 epochs with a
batch size of $4096$ takes roughly 10 days on a single
NVIDIA A100 GPU.\footnote{With an NVIDIA V100 GPU (16GB) training takes 16-18 days, and inference roughly a minute per event (rather than 20 seconds).}

\section{Effect of PSD}
\label{sec:effect-of-PSD}
\begin{figure}
  \includegraphics[width=\linewidth]{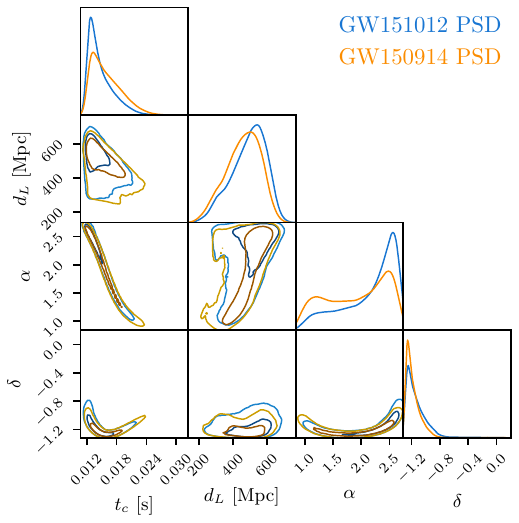}
  \caption{Comparison between DINGO evaluated on GW150914 using correct PSD as context, and using GW151012 PSD. These four parameters have a mean JSD of 0.020~nat.}
  \label{fig:wrong-PSD}
\end{figure}

To give a sense of the size of the effect of the PSD on the posterior,
we perform inference on GW150914, where we whiten the strain data with
the correct PSD, but provide instead the PSD for GW151012 as context
to the neural network. This gives a mean JSD across all parameters of
0.005~nat and a maximum JSD of 0.030~nat when compared against the
DINGO result using the correct PSD. 
Both of these numbers significantly exceed the JSDs between DINGO and
LALInference runs for all events [largest mean JSD: 0.001~nat (GW170729);
largest maximum JSD: 0.006~nat (GW170104, $m_1$)]. This demonstrates that
conditioning the neural network on the PSD is required at the level of
accuracy we achieve in this study.

The worst performing parameters with the incorrect PSD are $t_c$,
$d_L$, $\alpha$, and $\delta$, which have a mean JSD of 0.020~nat (see
Fig.~\ref{fig:wrong-PSD}). These results are consistent with the
expectation that the main effect due to using the wrong PSD is to
cause a change in the inferred amplitude of the signal in each
detector: $d_L$ becomes biased due to the overall amplitude being off,
whereas sky position and $t_c$ become biased from relative incorrect
amplitudes in the two detectors.

\section{Comparisons against standard samplers}

\begin{table}[t]
  \begin{ruledtabular}
    \begin{tabular}{c c c c c}
      Event & $\mathcal{M}~[\mathrm{M_\odot}]$ & $q$ & $\chi_{\mathrm{eff}}$ & $\chi_p$ \\
      \hline
      \multirow{2}{*}{GW150914} & $31.0_{-1.5}^{+1.5}$ & $0.85_{-0.21}^{+0.13}$ & $-0.03_{-0.12}^{+0.11}$ & $0.33_{-0.27}^{+0.40}$ \\
            & $31.1_{-1.5}^{+1.5}$ & $0.85_{-0.20}^{+0.13}$ & $-0.02_{-0.12}^{+0.11}$ & $0.33_{-0.27}^{+0.41}$ \\
      \cline{2-5}\multirow{2}{*}{GW151012} & $18.2_{-1.0}^{+1.1}$ & $0.60_{-0.32}^{+0.35}$ & $0.01_{-0.18}^{+0.20}$ & $0.30_{-0.23}^{+0.40}$ \\
            & $18.1_{-0.7}^{+0.8}$ & $0.63_{-0.34}^{+0.33}$ & $-0.00_{-0.15}^{+0.21}$ & $0.30_{-0.23}^{+0.40}$ \\
      \cline{2-5}\multirow{2}{*}{GW170104} & $25.5_{-1.8}^{+1.7}$ & $0.62_{-0.23}^{+0.33}$ & $-0.06_{-0.19}^{+0.16}$ & $0.39_{-0.28}^{+0.34}$ \\
            & $25.4_{-1.6}^{+1.6}$ & $0.63_{-0.22}^{+0.31}$ & $-0.07_{-0.17}^{+0.15}$ & $0.38_{-0.27}^{+0.34}$ \\
      \cline{2-5}\multirow{2}{*}{GW170729} & $49.2_{-8.1}^{+7.7}$ & $0.65_{-0.24}^{+0.30}$ & $0.25_{-0.26}^{+0.22}$ & $0.38_{-0.26}^{+0.33}$ \\
            & $49.6_{-8.2}^{+7.6}$ & $0.68_{-0.25}^{+0.28}$ & $0.27_{-0.27}^{+0.22}$ & $0.38_{-0.26}^{+0.32}$ \\
      \cline{2-5}\multirow{2}{*}{GW170809} & $29.8_{-1.9}^{+2.2}$ & $0.67_{-0.24}^{+0.29}$ & $0.06_{-0.16}^{+0.18}$ & $0.35_{-0.27}^{+0.38}$ \\
            & $29.9_{-1.8}^{+2.1}$ & $0.68_{-0.24}^{+0.28}$ & $0.07_{-0.15}^{+0.17}$ & $0.35_{-0.27}^{+0.38}$ \\
      \cline{2-5}\multirow{2}{*}{GW170814} & $27.2_{-1.2}^{+1.2}$ & $0.86_{-0.24}^{+0.13}$ & $0.08_{-0.12}^{+0.13}$ & $0.50_{-0.38}^{+0.31}$ \\
            & $27.1_{-1.1}^{+1.1}$ & $0.86_{-0.23}^{+0.13}$ & $0.08_{-0.11}^{+0.12}$ & $0.52_{-0.39}^{+0.29}$ \\
      \cline{2-5}\multirow{2}{*}{GW170818} & $32.7_{-2.8}^{+2.9}$ & $0.73_{-0.27}^{+0.24}$ & $-0.02_{-0.23}^{+0.21}$ & $0.51_{-0.35}^{+0.30}$ \\
            & $32.5_{-2.6}^{+2.7}$ & $0.74_{-0.27}^{+0.23}$ & $-0.05_{-0.22}^{+0.20}$ & $0.53_{-0.36}^{+0.28}$ \\
      \cline{2-5}\multirow{2}{*}{GW170823} & $38.9_{-4.1}^{+4.3}$ & $0.74_{-0.28}^{+0.23}$ & $0.06_{-0.20}^{+0.20}$ & $0.41_{-0.31}^{+0.36}$ \\
            & $38.9_{-3.9}^{+4.3}$ & $0.73_{-0.28}^{+0.24}$ & $0.06_{-0.20}^{+0.20}$ & $0.41_{-0.31}^{+0.36}$ \\
    \end{tabular}
    \caption{Comparison between DINGO (first line) and LALInference MCMC (second
      line) credible intervals. Median values and 90\% credible
      intervals are quoted.}
  \label{tab:intervals}
  \end{ruledtabular}
\end{table}

\begin{figure*}
   \subfloat[]{\includegraphics[width=\linewidth]{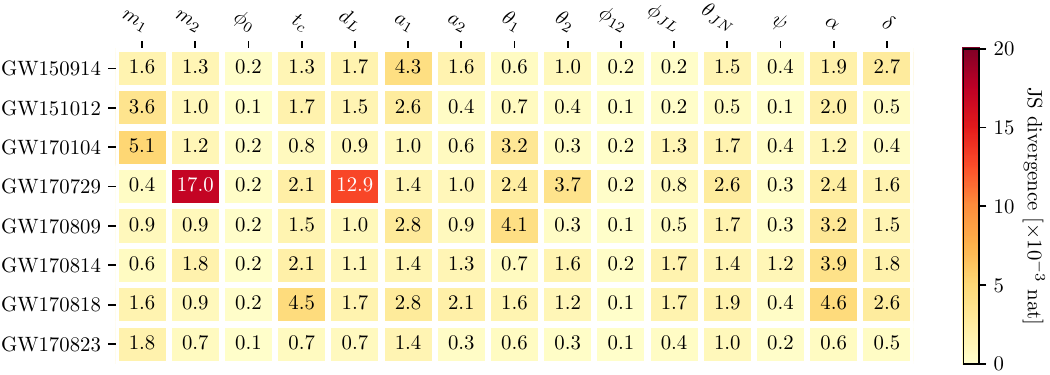}}
   \vspace{-5.0pt}

   \subfloat[]{\includegraphics[width=\linewidth]{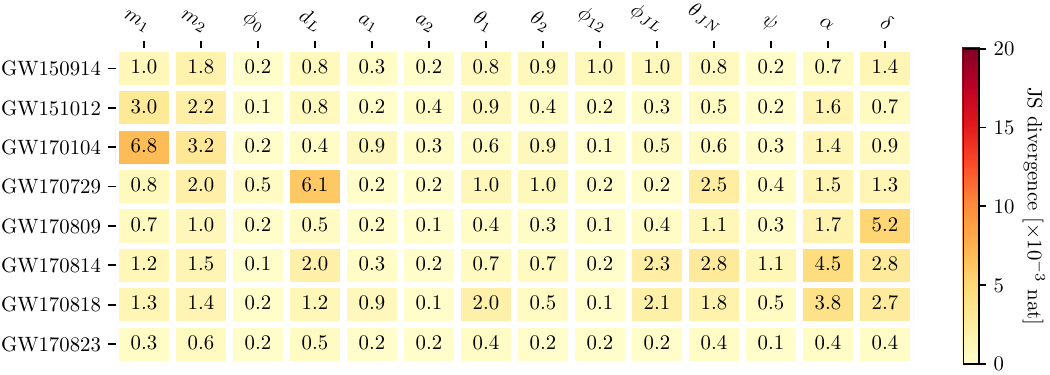}}
   \vspace{-5.0pt}
   \caption{JSDs between (a) DINGO and Bilby with time-distance-phase
     marginalization, and (b) DINGO and LALInference nested sampling
     with time marginalization. (Note that LALInference does not
     provide time posteriors when marginalization is used.) JSDs are
     calculated from 10,000 samples of each distribution, using
     Gaussian kernel density estimation. The mean values over 100
     different sample realizations are reported.  The mean JSDs over
     all events and parameters are 0.0015~nat and 0.0010~nat for (a)
     and (b), respectively. As mentioned in the main text, the average
     JSD between LALInference runs with identical settings but
     different random seeds is 0.0007~nat, which sets a
     \emph{practical} lower bound to the achievable values. Moreover,
     the DINGO framework allows us to obtain an arbitrary number of
     independent samples from the same distribution. This enables us
     to compute the JSDs between two sets of samples, that are, by
     construction, sampled from the same distribution. This value,
     0.0002~nat, provides a \emph{fundamental} lower bound for
     perfectly optimized samplers.}
  \label{fig:js-heatmap-detailed}
\end{figure*}

We compare our results against LALInference~\cite{Veitch:2014wba} with
MCMC and nested sampling algorithms, and with
Bilby~\cite{Ashton:2018jfp,Romero-Shaw:2020owr} with the
dynesty~\cite{Speagle_2020} nested sampling algorithm. To compare as
closely as possible with DINGO, we use the same data conditioning for
strain data and PSDs. However, we sample in chirp mass
$\mathcal{M} = (m_1m_2)^{3/5}/(m_1+m_2)^{1/5}$ and mass ratio
$q = m_2/m_1$ instead of component masses, since this simplifies the
form of the posterior and improves convergence. We also sample sky
position in the detector-based azimuth/zenith reference frame rather
than the $(\alpha,\delta)$ sky frame for Bilby. This also simplifies
the form of the posterior to improve
sampling~\cite{Romero-Shaw:2020owr}.

With standard samplers it is possible to analytically marginalize over
some parameters to reduce the dimensionality of the space. This
improves sampling performance, with the marginalized parameters
reconstructed in post-processing. For LALInference, we marginalize
over time of coalescence, and with Bilby we marginalize over time,
distance, and phase. Phase marginalization is only valid in the
absence of precession, but we had difficulty obtaining converged
results without it. For IMRPhenomPv2 waveforms, this should not lead
to a significant difference, but it should be kept in mind. For Bilby,
we use \verb|nlive=4000| and \verb|nact=50|.

We find closest agreement with LALInference MCMC, which is what we
report in the main text, and is displayed on all posterior plots. For
completeness we include in Fig.~\ref{fig:js-heatmap-detailed}
comparisons against LALInference nested sampling and Bilby with phase
marginalization. In general, when using phase marginalization, spin
and sky position are less well recovered, as were some events. To give
a sense of how JSD values translate into parameter estimates, we
provide 90\% credible intervals in Tab.~\ref{tab:intervals}, which are
all in extremely close agreement.

Deviations between posteriors obtained from DINGO and standard
samplers are associated with multiple sources of error. Firstly,
imperfect training of DINGO can lead to inaccurate results. Indeed,
training the neural networks is challenging due to the high
dimensionality of the input data to this inference problem, which
inspired us to adopt the GNPE method~\cite{Dax:2021myb} to improve
convergence. While the P--P plot in Fig.~2 of the main text suggests that
our networks are well converged, small deviations between posteriors
for real events could arise due to the networks not being fully
converged. Secondly, even the well-established standard samplers do
not produce perfect posterior samples. The mean JSD between
LALInference runs with identical settings but different random seeds
is a factor 3.5 higher than that expected for samples from identical
distributions (see Fig.~\ref{fig:js-heatmap-detailed} for
details). Thirdly, perfect agreement between (ideal implementations
of) DINGO and standard samplers is only expected for data consistent
with the training distribution. In reality, however, where the noise
is neither perfectly stationary nor Gaussian, the measured data is
slightly out-of-distribution. In those cases, there is no theoretical
guarantee that DINGO extrapolates to this data in the same way as
standard samplers. However, as stated in the conclusion, DINGO is not
limited to stationary Gaussian noise, and we plan to lift this
assumption in future work.

In the remainder of this section, we include 1D and 2D marginalized
posteriors for all nontrivial parameters for all events. All DINGO
posteriors are produced with 50,000 samples, except for the skymaps,
which use 10,000. LALInference posteriors use all samples produced
with the given sampler settings, typically of order 30,000--50,000.

\begin{figure*}[t]
  \stackinset{r}{}{c}{.95in}{\includegraphics[width=.44\textwidth]{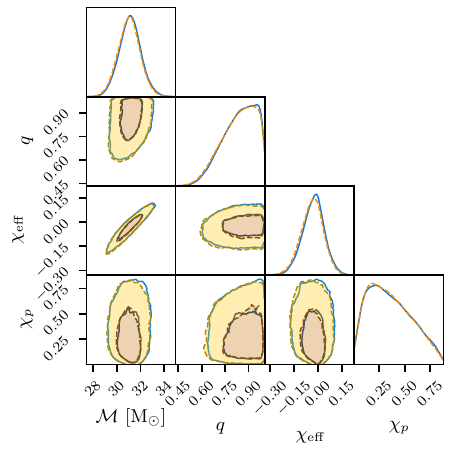}}{%
    \stackinset{c}{-.5in}{t}{-2in}{\includegraphics[width=0.4\textwidth]{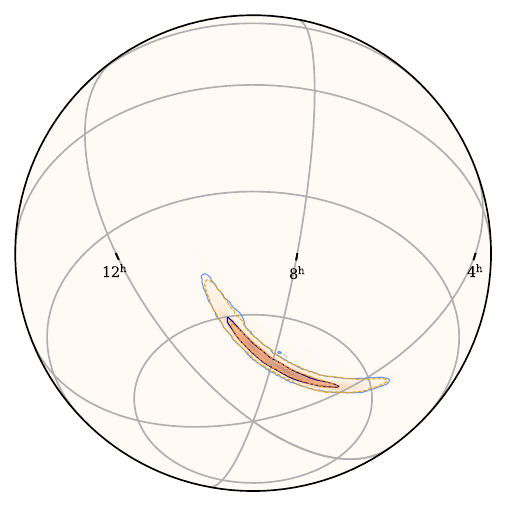}}{
      \includegraphics[width=\textwidth]{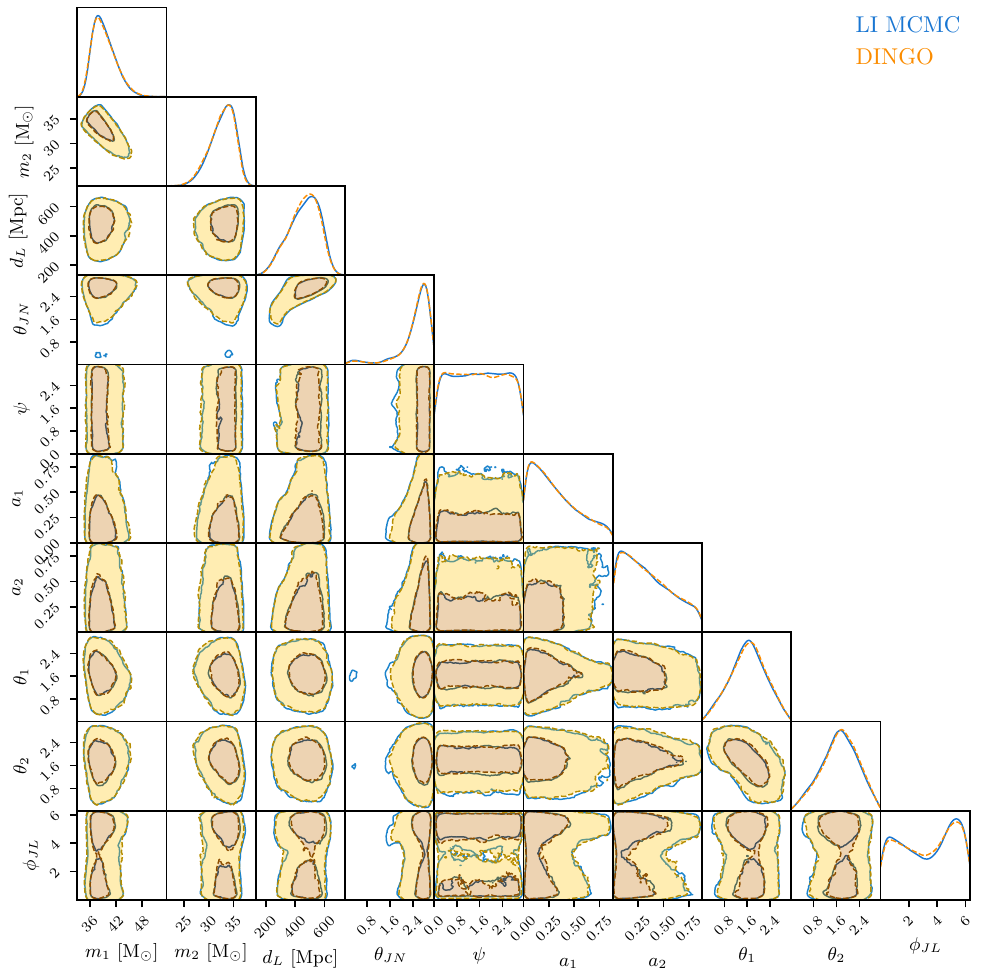}}}
  \cprotect\caption{\label{fig:posterior}Marginalized one- and two-
    dimensional posterior distributions for GW150914 over a subset of
    parameters, comparing DINGO (orange) and LALInference MCMC
    (blue). Relevant derived parameters plotted on right. Contours
    represent 50\% and 90\% credible regions. Posteriors reweighted to
    uniform source frame distance prior.}
\end{figure*}

\begin{figure*}
  \stackinset{r}{}{c}{.95in}{\includegraphics[width=.44\textwidth]{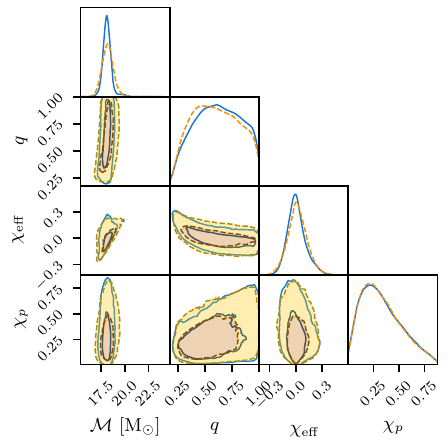}}{%
  \stackinset{c}{}{t}{-2in}{\includegraphics[width=0.6\textwidth]{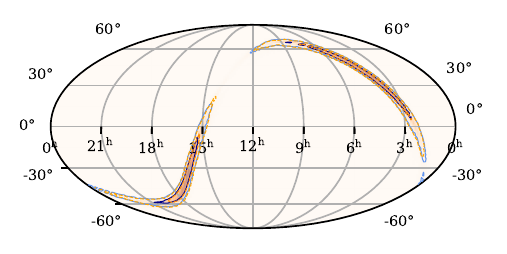}}{
    \includegraphics[width=\textwidth]{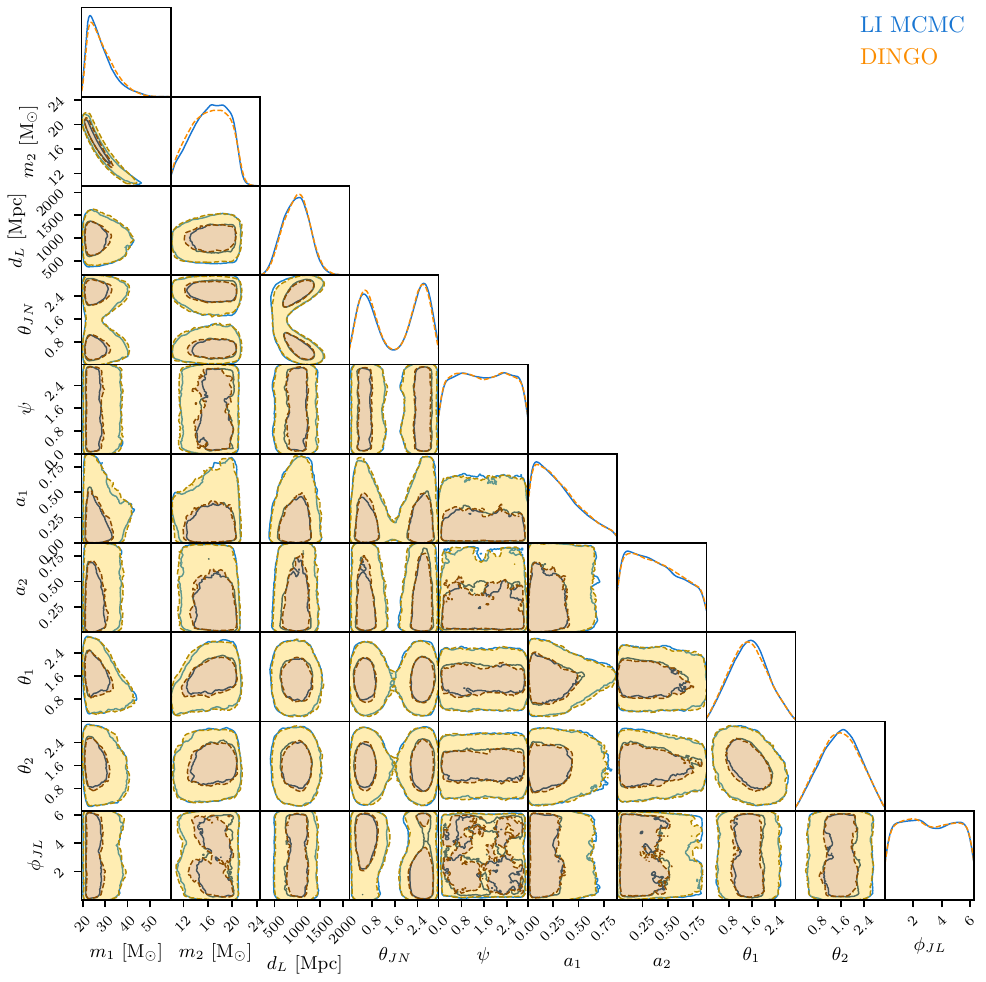}}}
  \cprotect\caption{\label{fig:posterior-GW151012}GW151012.}
\end{figure*}

\begin{figure*}
  \stackinset{r}{}{c}{.95in}{\includegraphics[width=.44\textwidth]{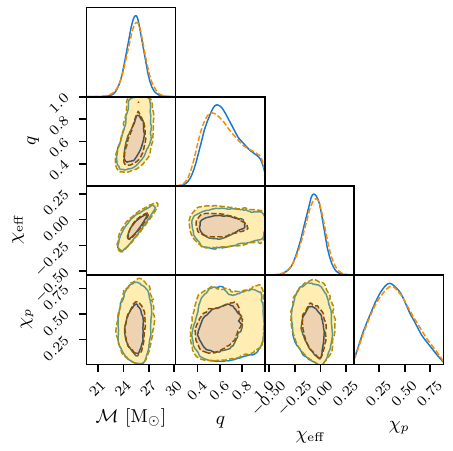}}{%
  \stackinset{c}{-.5in}{t}{-2in}{\includegraphics[width=0.4\textwidth]{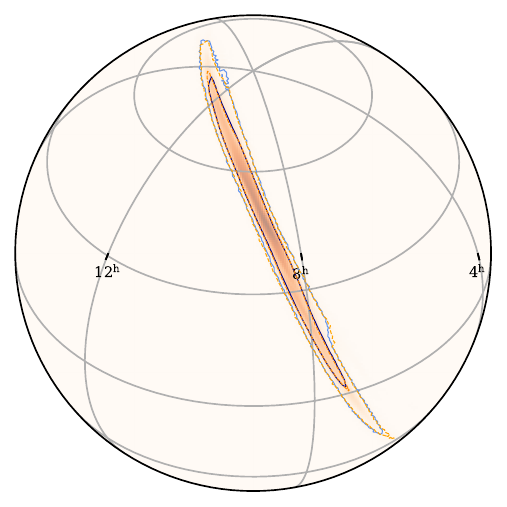}}{
    \includegraphics[width=\textwidth]{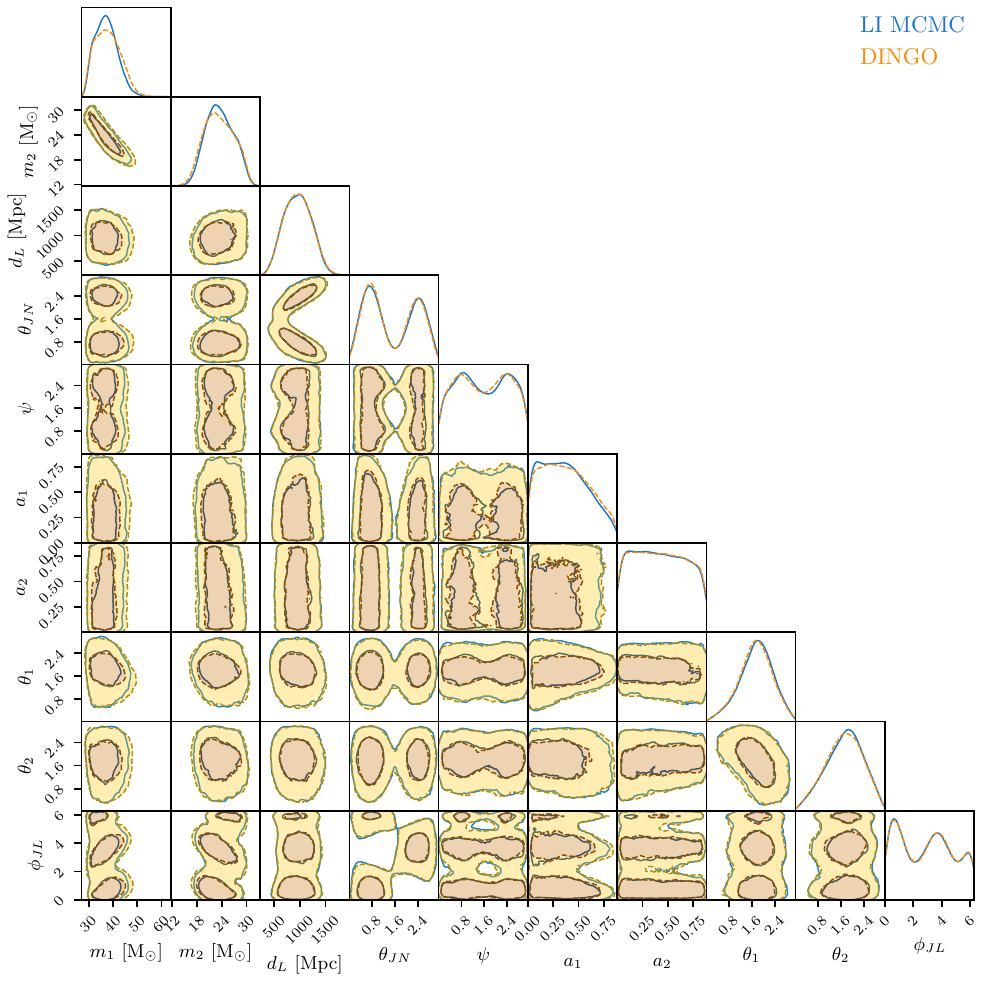}}}
  \cprotect\caption{\label{fig:posterior-GW170104}GW170104.}
\end{figure*}

\begin{figure*}
  \stackinset{r}{}{c}{.95in}{\includegraphics[width=.44\textwidth]{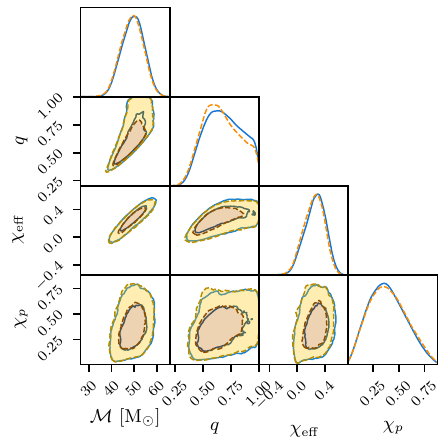}}{%
  \stackinset{c}{-.5in}{t}{-2in}{\includegraphics[width=0.4\textwidth]{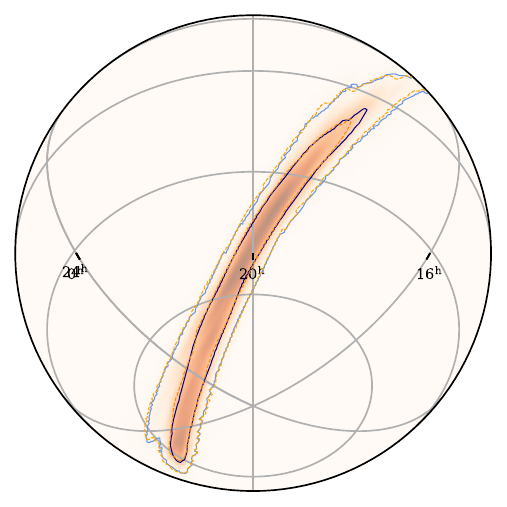}}{
    \includegraphics[width=\textwidth]{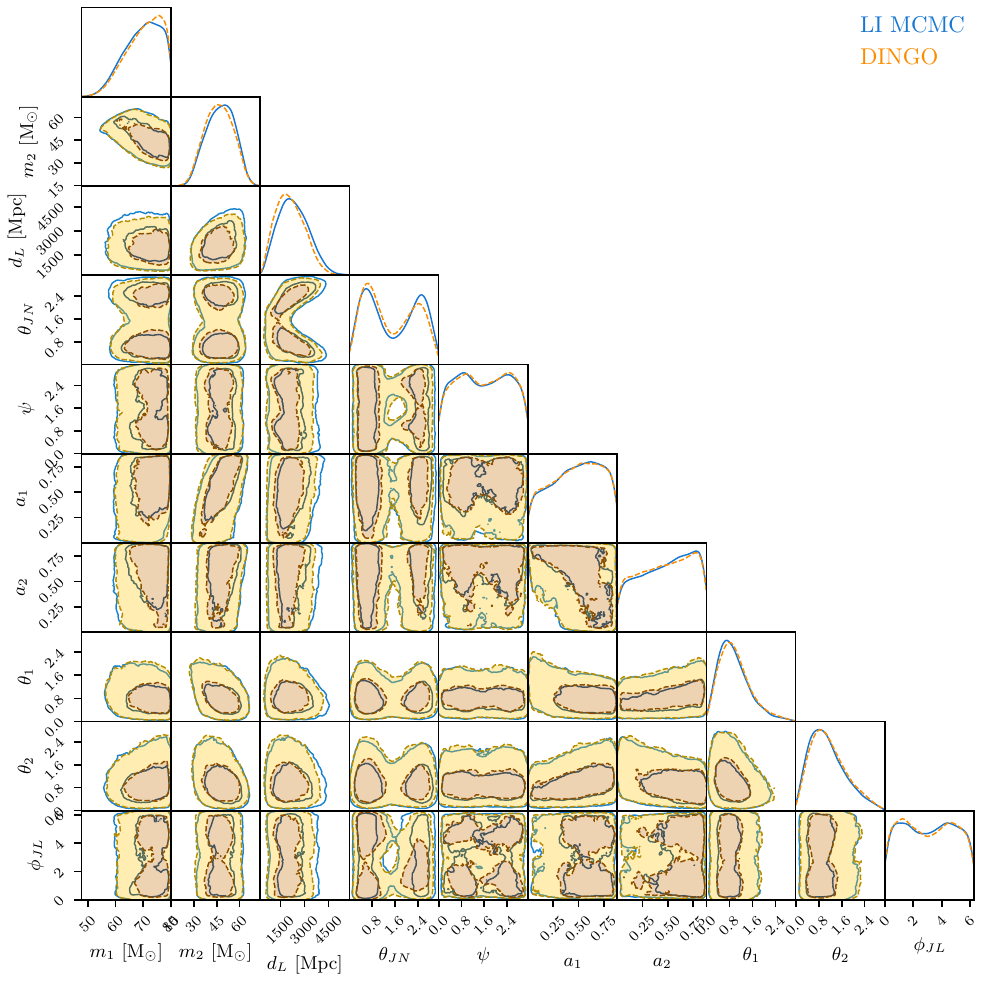}}}
  \cprotect\caption{\label{fig:posterior-GW170729}GW170729.}
\end{figure*}

\begin{figure*}
  \stackinset{r}{}{c}{.95in}{\includegraphics[width=.44\textwidth]{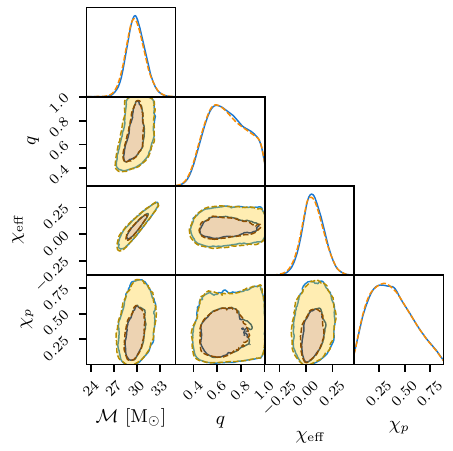}}{%
  \stackinset{c}{-.5in}{t}{-2in}{\includegraphics[width=0.4\textwidth]{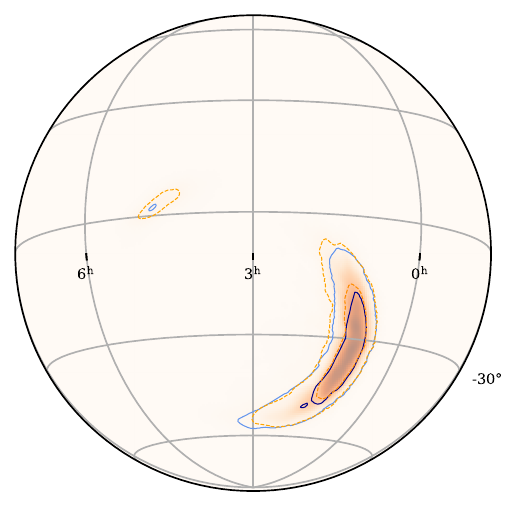}}{
    \includegraphics[width=\textwidth]{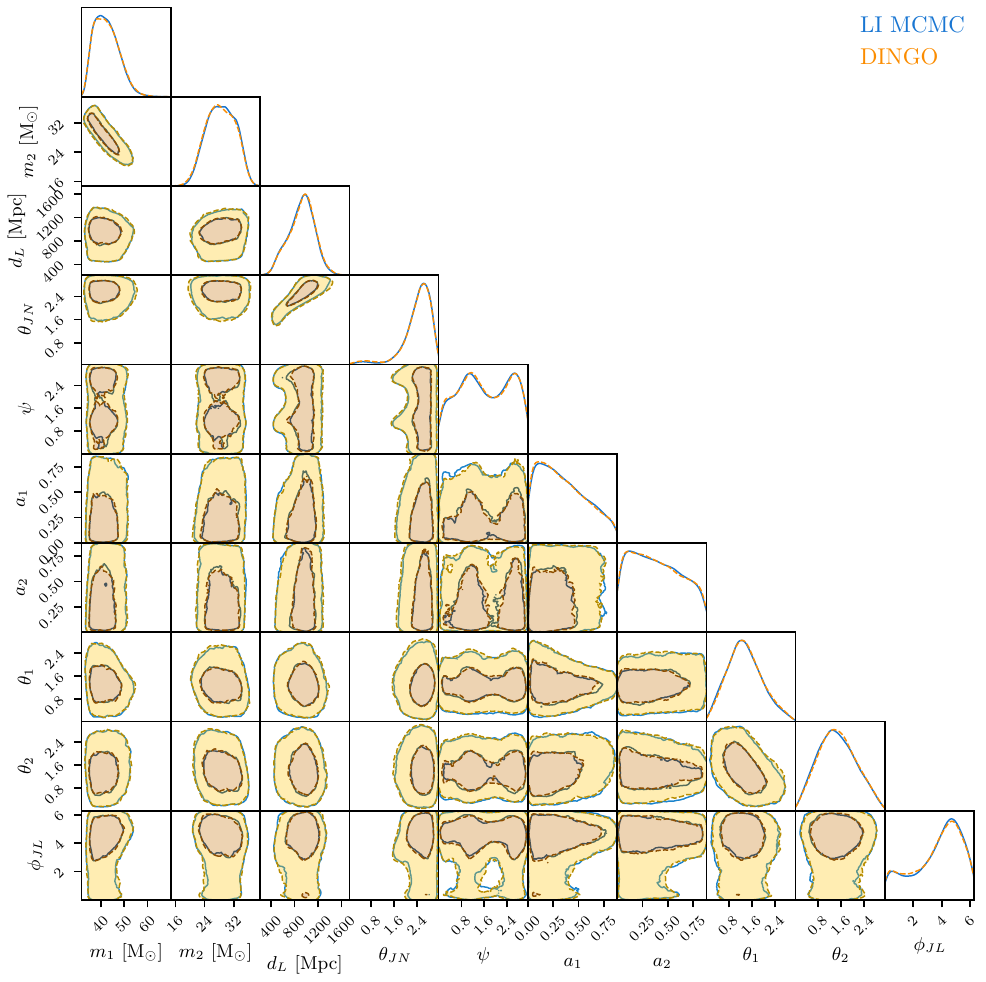}}}
  \cprotect\caption{\label{fig:posterior-GW170809}GW170809.}
\end{figure*}

\begin{figure*}
  \stackinset{r}{}{c}{.95in}{\includegraphics[width=.44\textwidth]{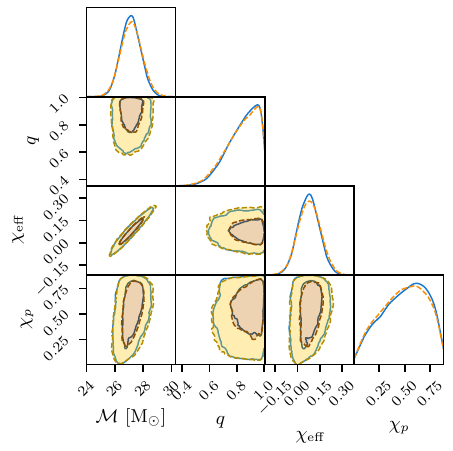}}{%
  \stackinset{c}{-.5in}{t}{-2in}{\includegraphics[width=0.4\textwidth]{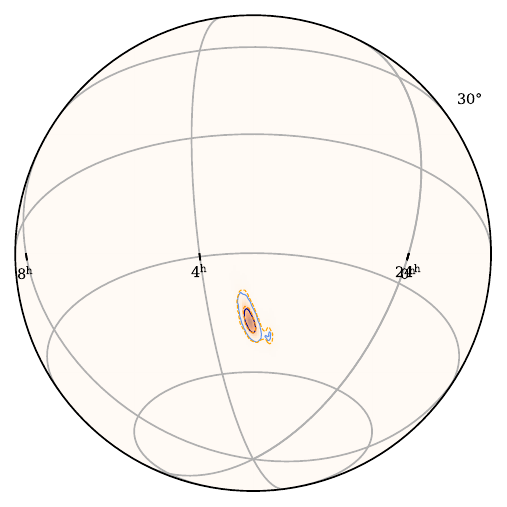}}{
    \includegraphics[width=\textwidth]{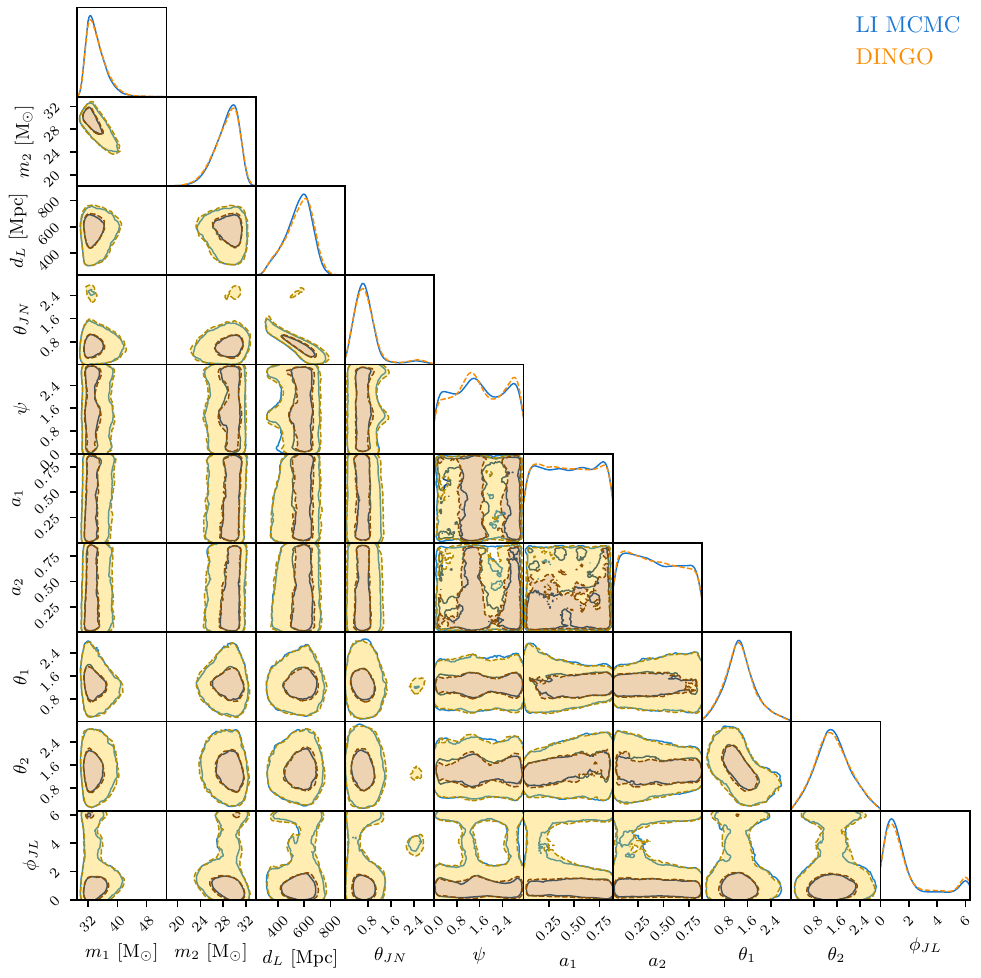}}}
  \cprotect\caption{\label{fig:posterior-GW170814}GW170814. This is
    the only event analyzed as a three-detector event.}
\end{figure*}

\begin{figure*}
  \stackinset{r}{}{c}{.95in}{\includegraphics[width=.44\textwidth]{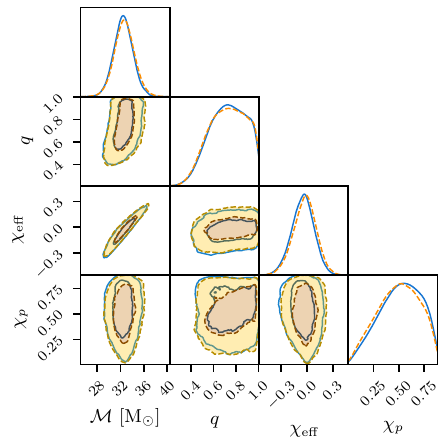}}{%
  \stackinset{c}{-.5in}{t}{-2in}{\includegraphics[width=0.4\textwidth]{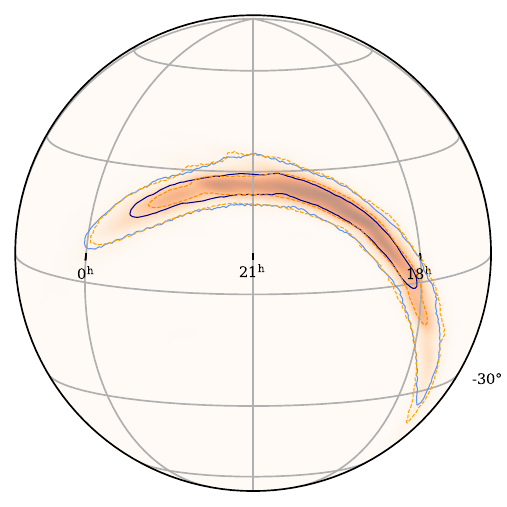}}{
    \includegraphics[width=\textwidth]{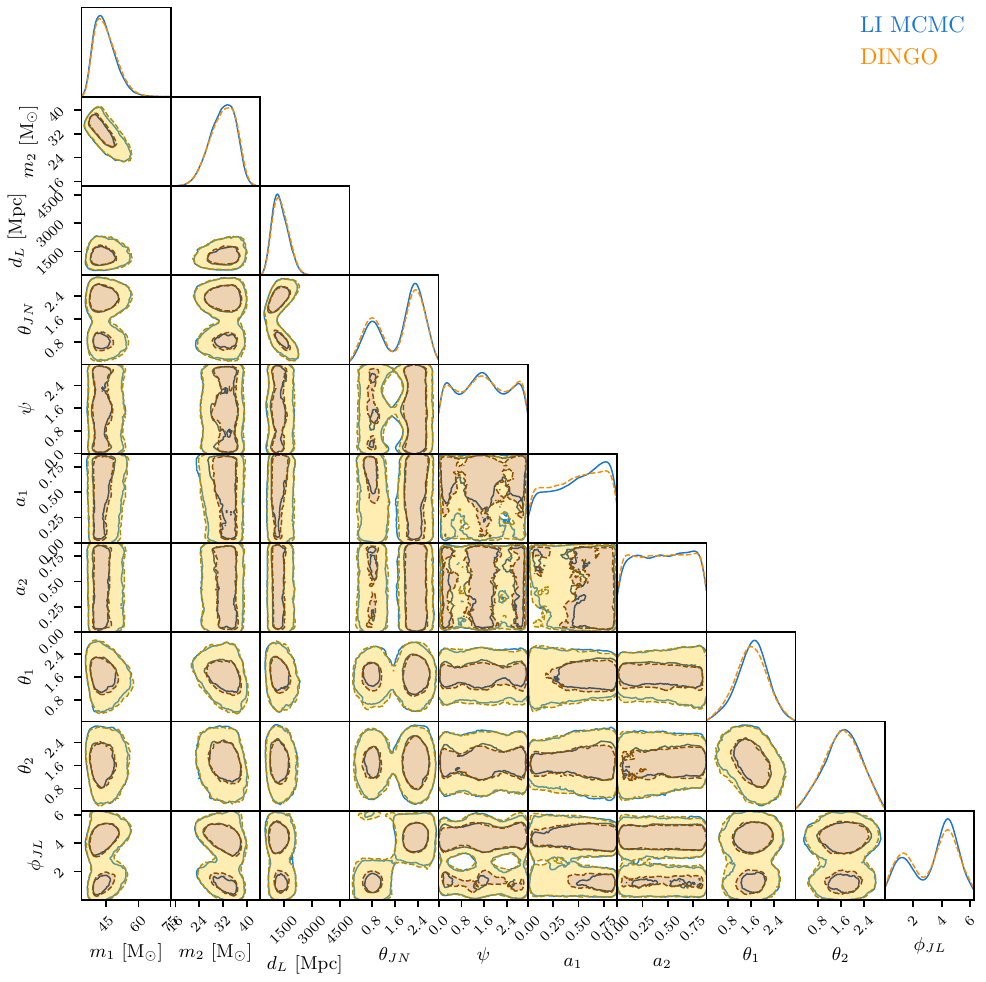}}}
  \cprotect\caption{\label{fig:posterior-GW170818}GW170818.}
\end{figure*}

\begin{figure*}
  \stackinset{r}{}{c}{.95in}{\includegraphics[width=.44\textwidth]{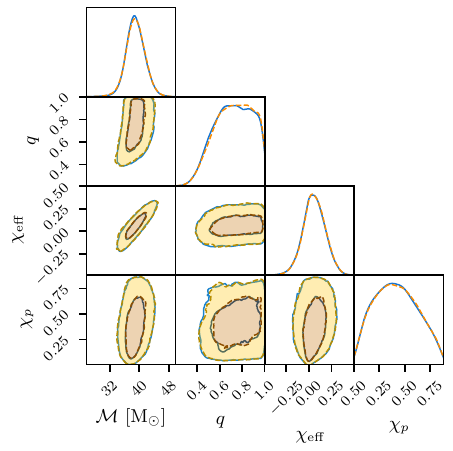}}{%
  \stackinset{c}{}{t}{-2in}{\includegraphics[width=0.6\textwidth]{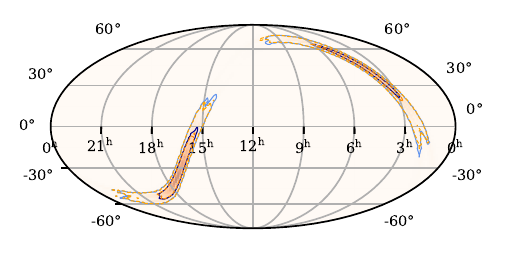}}{
    \includegraphics[width=\textwidth]{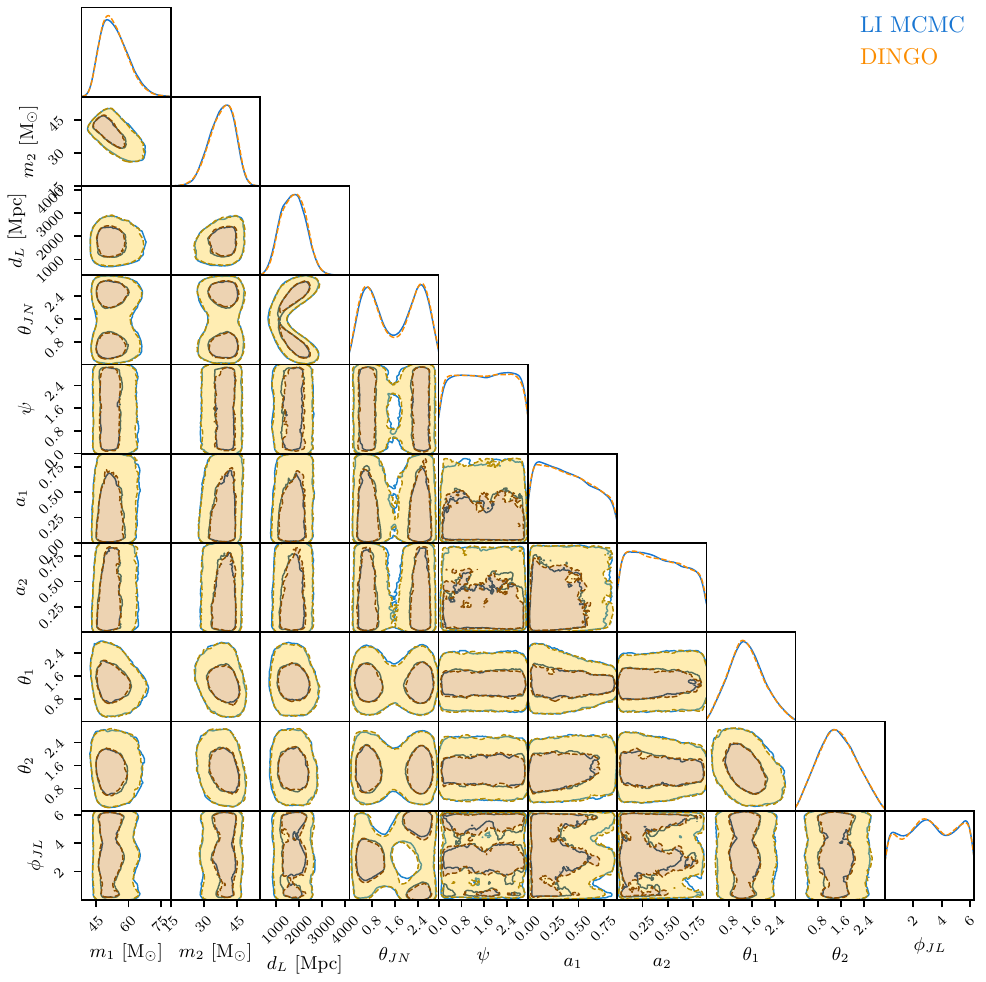}}}
  \cprotect\caption{\label{fig:posterior-GW170823}GW170823.}
\end{figure*}

\end{document}